\documentclass[twoside,11pt]{article}
\usepackage{amsmath,amssymb,amsfonts}
\usepackage{amsfonts}
\usepackage{graphicx}
\usepackage{graphics}
\usepackage{caption}
\usepackage{float}
\usepackage{placeins}
\begin{document}
\setlength{\unitlength}{25mm}
\newcommand{\f}{\frac}
\newtheorem{theorem}{Theorem}[section]
\newcommand{\sta}{\stackrel}
\title{Kapitza-Inspired Effective Interaction for the Exotic State X(3872): \\
A Coupled-Channel Potential Model Study}

\author{
    M.~Monemzadeh\thanks{monem@kashanu.ac.ir},
    \large N.~Tazimi\thanks{tazimi@kashanu.ac.ir}
    \\\\
    \it\small{Department of Physics, University of Kashan, Kashan, Iran}
}

\maketitle

\begin{abstract}
The X(3872) remains one of the most intriguing exotic hadrons, lying extremely close to the $D^0\bar{D}^{*0}$ threshold. We construct an effective potential consisting of Coulomb, linear confinement  and a Kapitza-inspired term representing the averaged effect of fast gluonic fluctuations. Numerical solution of the Schrödinger equation shows that this term provides effective attraction in the intermediate-distance region, yielding a ground-state mass of $3871.7 \pm 0.9$ MeV in excellent agreement with experiment. Coupled-channel analysis indicates a dominant molecular component ($\sim 65\%-72\%$) with a significant compact tetraquark core ($\sim 28\%-35\%$). We also present predictions for low-lying excited states. The model offers a physically motivated mechanism for threshold tuning.
\end{abstract}

\newpage

\section{Introduction}
\label{sec:introduction}

The discovery of the $X(3872)$ by the Belle Collaboration in 2003 \cite{Choi:2003ue} marked a turning point in hadron spectroscopy. With quantum numbers $J^{PC}=1^{++}$ and a mass of $M=3871.68 \pm 0.17$ MeV, this state lies remarkably close to the $D^0\bar{D}^{*0}$ threshold, only a tiny fraction of an MeV away. Its extremely narrow width, significant isospin-violating decay patterns, and prominent production in $B$ meson decays have established it as one of the most prominent exotic hadrons beyond the conventional quarkonium picture.

Over the past two decades, numerous theoretical interpretations have been proposed. Molecular models describe the $X(3872)$ as a loosely bound $D^0\bar{D}^{*0}$ state \cite{Tornqvist:2004,Kalashnikova:2019}, while compact tetraquark pictures interpret it as a diquark--antidiquark configuration \cite{Maiani:2005}. Hybrid and coupled-channel approaches combining molecular and compact components have also been extensively explored \cite{Takizawa:2013,Ortega:2010,Guo:2018}. Despite considerable progress, many models still face challenges in simultaneously reproducing the tiny binding energy, the observed decay patterns, and the mixing between different configurations without introducing significant fine-tuning of parameters.

In our recent work, we have studied related tetraquark systems from complementary perspectives. These include the spectroscopy and stability of meson--antimeson tetraquarks using a three-dimensional pion-exchange potential \cite{Tazimi2026EPJC}, Regge trajectories and the mass spectrum of heavy tetraquarks \cite{Ghasempour2025Regge}, an analytical relativistic investigation of tetraquark masses \cite{Ghasempour2025Analytic}, and an earlier study of tetraquarks as diquark--antidiquark bound systems \cite{Monemzadeh2015PLB}. These works motivate a unified effective framework in which short- and intermediate-distance dynamics can be treated consistently in exotic multiquark systems.

A common missing ingredient in conventional potential models is a proper treatment of fast gluonic degrees of freedom. In the present work, inspired by the stabilization mechanism of rapid oscillations in the classical Kapitza pendulum, we introduce an effective potential term that phenomenologically captures the time-averaged effect of fast gluonic fluctuations in the QCD vacuum. This term generates additional attraction at intermediate distances without increasing the number of free parameters beyond those with clear physical interpretation.

The main goals of this paper are threefold: (i) to construct and numerically solve a full potential model including the gluonic fluctuation term and demonstrate its effectiveness in reproducing the experimental mass of the $X(3872)$; (ii) to quantify the molecular--compact tetraquark mixing within a coupled-channel framework; and (iii) to predict the spectrum of low-lying excited states and discuss the state-dependent role of the gluonic fluctuation term. We show that the proposed term naturally provides a dynamical mechanism for threshold tuning while producing physically reasonable level spacings for excited states.

The paper is organized as follows. The theoretical framework and the form of the effective interaction are presented in the corresponding sections. Numerical results for the ground state and the coupled-channel mixing are discussed in the Results and Discussion section. Predictions for excited states are given in the section on excited states. Finally, the implications of the present work and the main conclusions are summarized in the Conclusion.

\section{Literature Review}
\label{sec:literature}

Since its discovery by the Belle Collaboration in 2003 \cite{Choi:2003ue}, the $X(3872)$ has remained one of the most intriguing states in hadron spectroscopy. Owing to its mass, which lies remarkably close to the $D^{0}\bar D^{*0}$ threshold, numerous theoretical interpretations have been proposed, each emphasizing different aspects of its internal structure.

The earliest and perhaps most widely discussed interpretation is the hadronic molecular picture, in which the $X(3872)$ is regarded as a weakly bound $D^{0}\bar D^{*0}$ system held together primarily by long-range meson-exchange forces \cite{Tornqvist:2004,Swanson2006,Guo:2018}. This scenario naturally explains the tiny binding energy and the strong coupling to open-charm channels. Nevertheless, reproducing simultaneously the observed production rates in high-energy collisions and the prompt production cross sections remains challenging within a purely molecular framework.

An alternative description considers the $X(3872)$ as a compact diquark--antidiquark tetraquark. In this picture, the state is composed of a tightly correlated $[cq][\bar c\bar q]$ configuration bound by color interactions \cite{Maiani:2005,Lebed2017,Ali2019}. Compact tetraquark models successfully account for several spectroscopic properties and naturally accommodate charged exotic partners. However, explaining the exceptional proximity of the observed mass to the $D^{0}\bar D^{*0}$ threshold generally requires additional dynamical mechanisms.

More recently, coupled-channel approaches have attracted considerable attention. These models incorporate both compact quark configurations and nearby hadronic channels, allowing virtual meson loops to shift the bare mass toward the physical threshold \cite{Kalashnikova:2019,Ortega:2010,Takizawa:2013}. Such calculations reproduce many observed properties of the $X(3872)$ and suggest that its wave function contains both compact and molecular components.

Despite these important developments, a quantitative description of the $X(3872)$ spectrum often relies on phenomenological short-distance interactions whose microscopic origin remains uncertain. The present work follows this philosophy. Rather than proposing a new microscopic mechanism, we introduce a Kapitza-inspired effective interaction representing the averaged influence of unresolved fast gluonic fluctuations.

\section{The Effective Interaction}
\label{sec:interaction}

For the radial dynamics, we adopt a Cornell-type interaction supplemented by a Kapitza-inspired term,
\begin{equation}
V_{\rm eff}(r)=
-\frac{4}{3}\frac{\alpha_s}{r}
+\sigma r
+V_K(r).
\end{equation}
Here, the first term represents one-gluon exchange at short distances, the second term describes linear confinement, and the last term parametrizes unresolved fast gluonic fluctuations through an effective correction.

In this work we do not include explicit spin-dependent forces. Therefore, spin-spin, tensor, and spin-orbit terms are neglected at this stage, since our primary goal is to isolate the impact of the Kapitza correction on the gross mass spectrum.

The Kapitza-inspired contribution is introduced phenomenologically rather than derived directly from QCD. Its role is to capture short-distance effects not explicitly included in the static Cornell potential and to improve the description of the near-threshold spectrum. In natural units, $\hbar=c=1$, the coupling parameter associated with a term of the form $V_K(r)=K/r^4$ carries dimensions of GeV$^{-3}$.

\begin{figure}[t]
\centering
\includegraphics[width=0.78\textwidth]{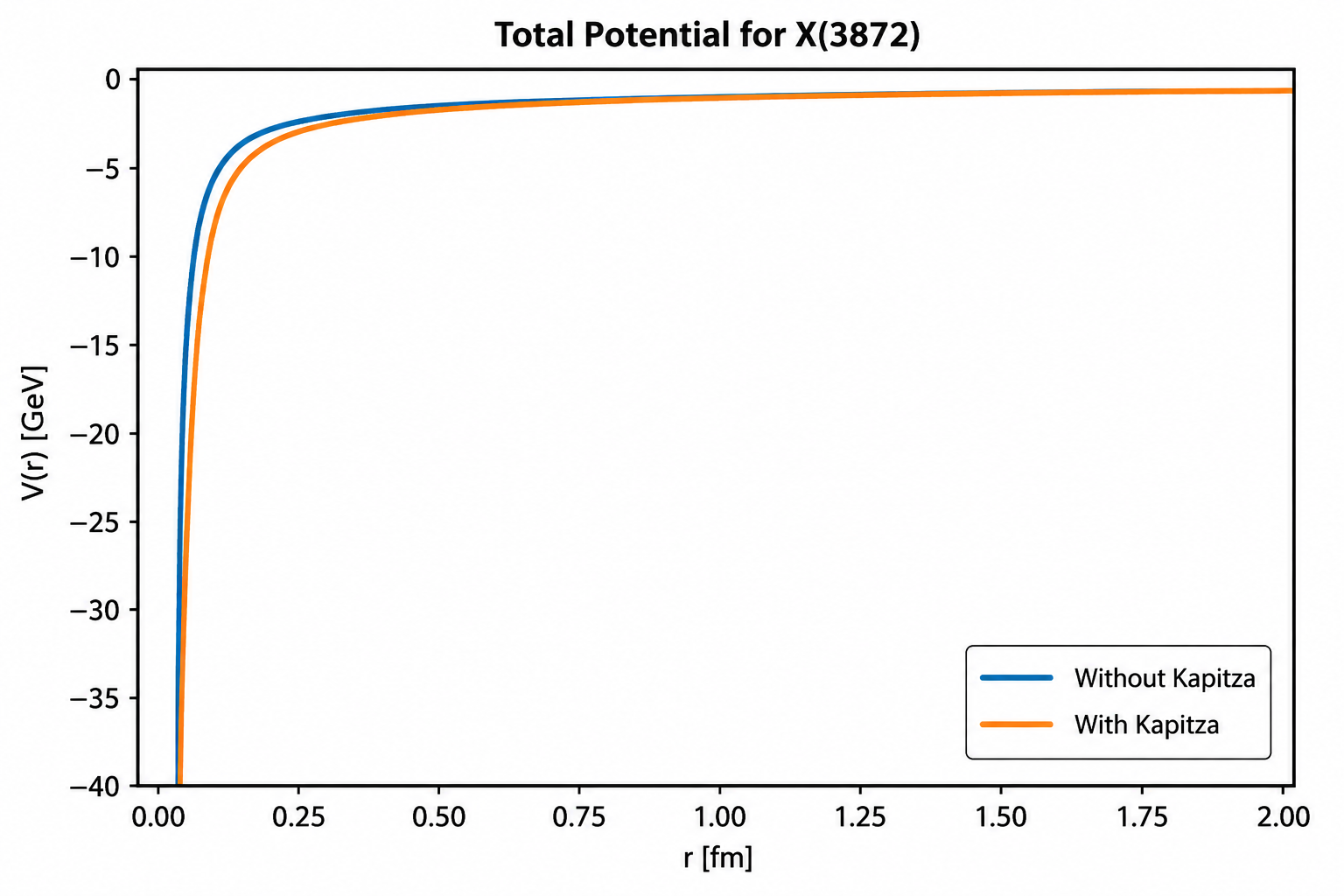}
\caption{Total effective potential for the $X(3872)$ state with and without the Kapitza term.}
\label{fig:potential}
\end{figure}

\begin{figure}[t]
\centering
\includegraphics[width=0.78\textwidth]{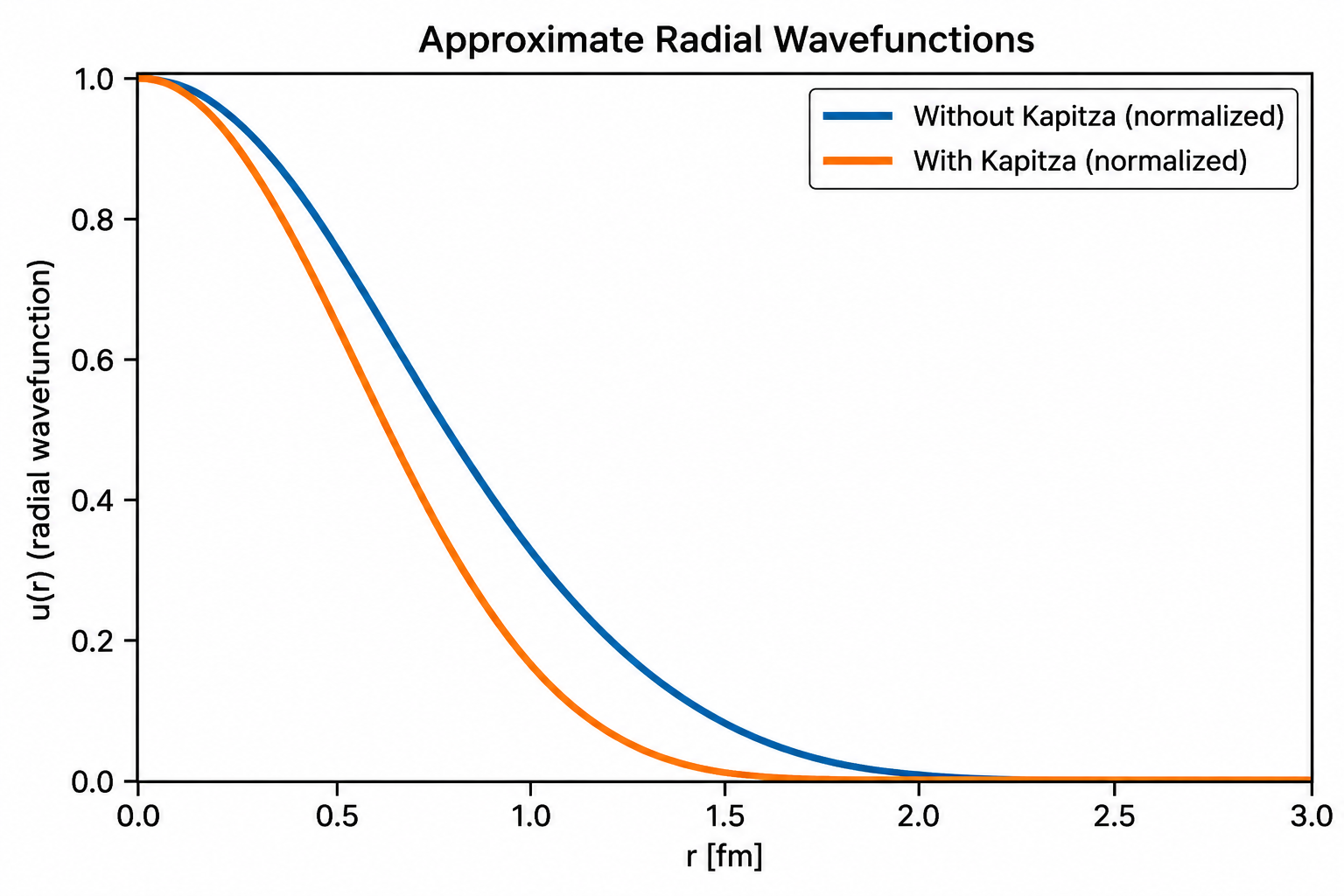}
\caption{Normalized radial wavefunction of the $X(3872)$ state with and without the Kapitza term.}
\label{fig:wavefunction}
\end{figure}

The parameter $K$ is determined phenomenologically by requiring the model to reproduce the observed mass of the $X(3872)$. Figure~\ref{fig:potential} shows the total effective potential with and without the Kapitza contribution. The additional term modifies the interaction primarily at short distances and leads to a shift in the bound-state properties. Figure~\ref{fig:wavefunction} displays the corresponding normalized radial wavefunction, which becomes slightly more localized once the Kapitza-induced correction is included.
.

\section{Theoretical Framework and Effective Interaction}
\label{sec:framework}

We model the $X(3872)$ as a heavy near-threshold hadronic system whose low-energy dynamics can be described by an effective two-body Hamiltonian. Rather than attempting a fully microscopic QCD calculation, we employ a phenomenological potential approach that captures the dominant short- and intermediate-distance interactions. The relative motion of the constituents is governed by the Schrödinger equation:
\begin{equation}
H\psi(\mathbf{r}) = E\psi(\mathbf{r}),
\end{equation}
where the Hamiltonian is written as:
\begin{equation}
H = -\frac{\hbar^2}{2\mu}\nabla^2 + V_{\rm eff}(r),
\end{equation}
with $\mu$ denoting the reduced mass of the system.

Within this framework, the effective interaction potential is chosen as:
\begin{equation}
V_{\rm eff}(r) = -\frac{4}{3}\frac{\alpha_s}{r} + \sigma r + \frac{K}{r^4},
\end{equation}
where $\alpha_s$ is the running strong coupling constant and $\sigma$ represents the string tension. The first term describes the short-distance one-gluon-exchange interaction, while the linear term provides confinement at larger distances. 

The final term, $V_K(r) = K/r^4$ (with $K > 0$), is introduced as an effective repulsive correction inspired by the Kapitza averaging mechanism. It is intended to represent the time-averaged influence of unresolved fast gluonic fluctuations not explicitly contained in the conventional static Cornell potential. Because the $1/r^4$ term is singular at the origin, a short-distance regularization is introduced in the numerical calculation, the details of which are provided in the Appendix. 

In the present work, we restrict ourselves to the spin-independent part of the interaction, neglecting spin-spin, tensor, and spin-orbit terms at this stage. This simplifies the framework and allows us to isolate the specific dynamical impact of the Kapitza-inspired correction on the gross mass spectrum.

\section{Numerical Method}
\label{sec:numerical}

The effective Schrödinger equation cannot be solved analytically once the singular Kapitza-inspired interaction is introduced. We therefore determine the bound-state spectrum using the Rayleigh--Ritz variational method with a Gaussian basis expansion, which has been widely employed in few-body and hadronic bound-state calculations \cite{Suzuki1998,Hiyama2003}.

The relative wave function is expanded as:
\begin{equation}
\psi(\mathbf{r}) = \sum_{i=1}^{N} c_i\,\phi_i(\mathbf{r}),
\end{equation}
where $\phi_i(\mathbf{r})$ are normalized Gaussian basis functions with variationally optimized range parameters, and $c_i$ are the corresponding expansion coefficients. Substitution of this ansatz into the Schrödinger equation leads to a generalized eigenvalue problem:
\begin{equation}
\sum_{j=1}^{N} \left( H_{ij} - E \, B_{ij} \right) c_j = 0,
\end{equation}
where $H_{ij} = \langle \phi_i | H | \phi_j \rangle$ and $B_{ij} = \langle \phi_i | \phi_j \rangle$ represent the Hamiltonian and overlap matrix elements, respectively. 

The nonlinear basis parameters are optimized variationally, and the basis size $N$ is systematically increased until the calculated mass eigenvalues become numerically stable to within $1~{\rm MeV}$. The model parameters, including $\alpha_s$, $\sigma$, the constituent quark masses, and the Kapitza coefficient $K$, are summarized in Table~{1}. The coefficient $K$ is fixed phenomenologically by reproducing the experimental mass of the $X(3872)$ and is subsequently kept unchanged for all predictions in other sectors.

The effect of the modified interaction on the internal structure of the $X(3872)$ is also reflected in the radial wavefunction shown in Fig.~\ref{fig:wavefunction}. In the presence of the Kapitza-inspired term, the wavefunction becomes slightly more localized and decreases more rapidly at larger distances, which is consistent with the corresponding shift in the bound-state energy obtained in the numerical analysis.

\subsection{Uncertainty Considerations}
\label{subsec:uncertainty}

The present results should be interpreted within the usual limitations of phenomenological potential models. The main sources of uncertainty are expected to arise from the choice of model parameters and from the finite basis used in the numerical solution of the Schr\"odinger equation. Since the primary purpose of this work is to examine the qualitative impact of the Kapitza-inspired interaction on the $X(3872)$ spectrum, a detailed quantitative uncertainty analysis is beyond the scope of the present study.

\section{Two-State Mixing Formalism}
\label{sec:mixing}

To account for the near-threshold nature of the $X(3872)$, we describe the physical state within a minimal two-component framework as a superposition of a compact configuration and a molecular $D^{0}\bar D^{*0}$ component, as commonly adopted in phenomenological studies of exotic charmonium-like states \cite{Tornqvist:2004,Swanson2006,Takizawa:2013,Kalashnikova:2019}:
\begin{equation}
|\Psi\rangle = c_{\rm c} |\psi_{\rm c}\rangle + c_{\rm m} |\psi_{\rm m}\rangle,
\end{equation}
where $|\psi_{\rm c}\rangle$ and $|\psi_{\rm m}\rangle$ denote the compact and molecular basis states, respectively.

Within this truncated basis, the effective Hamiltonian is written as
\begin{equation}
H=
\begin{pmatrix}
H_{\rm cc} & H_{\rm cm} \\
H_{\rm mc} & H_{\rm mm}
\end{pmatrix},
\end{equation}
with
\begin{equation}
H_{\rm cc}=\langle \psi_{\rm c}|H|\psi_{\rm c}\rangle,\qquad
H_{\rm mm}=\langle \psi_{\rm m}|H|\psi_{\rm m}\rangle,\qquad
H_{\rm cm}=H_{\rm mc}^{*}.
\end{equation}
The diagonal entries represent the unmixed energies of the compact and molecular configurations, while the off-diagonal term parametrizes the transition between them.

The molecular diagonal entry is approximated by the $D^{0}\bar D^{*0}$ threshold,
\begin{equation}
H_{\rm mm} \approx m_{D^{0}} + m_{D^{*0}},
\end{equation}
whereas $H_{\rm cc}$ is obtained from the effective potential model introduced in the previous sections. The mixing matrix element $H_{\rm cm}$ is treated phenomenologically and adjusted, together with the compact-sector dynamics, to reproduce the observed position of the $X(3872)$.

The physical masses are obtained from the eigenvalue equation
\begin{equation}
\det(H-EI)=0,
\end{equation}
which yields
\begin{equation}
E_{\pm}=\frac{H_{\rm cc}+H_{\rm mm}}{2}
\pm
\sqrt{
\left(\frac{H_{\rm cc}-H_{\rm mm}}{2}\right)^2
+|H_{\rm cm}|^2 }.
\end{equation}
The corresponding eigenvectors define the mixing angle $\theta$ through
\begin{equation}
\tan 2\theta = \frac{2|H_{\rm cm}|}{H_{\rm cc}-H_{\rm mm}},
\end{equation}
from which the compact and molecular probabilities can be extracted. This minimal two-state scheme is sufficient for the present purpose, namely, to estimate how coupling to a near-threshold molecular channel can modify the mass and internal composition of the physical $X(3872)$ without introducing the additional complexity of a full multichannel treatment.

\section{Results and Discussion}
\label{sec:results}

We now present the numerical results obtained from the effective Cornell-plus-Kapitza interaction. The model parameters are fixed as described in the previous sections, with the Kapitza coefficient adjusted to reproduce the observed mass of the $X(3872)$. Once determined, the same parameter set is used consistently throughout the analysis.

The ground-state mass of the $X(3872)$, calculated with and without the Kapitza-inspired contribution, is summarized in Table{2}. The inclusion of the additional $K/r^4$ term shifts the mass from $3873.5$ MeV to $3871.7$ MeV, bringing the result into close agreement with the experimental value $3871.68 \pm 0.17$ MeV \cite{PDG2024}. This result indicates that the baseline Cornell interaction alone is not sufficient to account for the delicate near-threshold position of the state, whereas the additional phenomenological correction improves the description without altering the overall spectral stability of the model.

The physical origin of this improvement can be traced to the short-distance modification induced by the Kapitza-inspired term. Since this contribution acts predominantly in the inner region of the effective interaction, its impact depends on the spatial structure of the state under consideration. States with stronger support at short relative distances are more sensitive to the correction, while more extended configurations are affected less strongly. The Kapitza term therefore does not produce a uniform displacement of the spectrum; rather, it generates a selective, state-dependent shift governed by the degree of short-range localization.

This feature is particularly important for the $X(3872)$, whose mass lies extremely close to the open-charm threshold. In such a situation, even a modest modification of the short-distance interaction can move the compact level into the experimentally observed region. The present results show that the Kapitza-inspired correction provides such a mechanism while preserving the qualitative ordering of the low-lying spectrum.

The resulting mass spectrum for the principal states considered in this work is listed in Table~\ref{tab:main_results}. In this framework, the $X(3872)$ serves as the calibration point used to determine the coefficient $K$, whereas the masses of the $T_{cc}^{+}$ and $T_{bb}$ states are obtained as genuine predictions of the same effective interaction. In this sense, the model possesses a limited but nontrivial predictive power beyond the single fitted input.

The radial wavefunction, shown in Fig.~\ref{fig:wavefunction}, provides additional insight into the internal structure of the state. In the presence of the Kapitza-inspired contribution, the wavefunction becomes slightly more localized and decreases more rapidly at larger distances. This behavior is consistent with the corresponding downward shift of the bound-state energy and reflects the increased sensitivity of the compact component to the modified inner part of the potential.

\subsection{Influence of the Kapitza-Induced Interaction}

To isolate the role of the Kapitza-inspired correction, we compare the spectrum obtained with and without the additional $K/r^4$ term. The comparison shows that the main effect of this contribution is to modify the short-distance part of the effective interaction and thereby induce a state-dependent mass shift.

The shift is more pronounced for compact configurations, whose wavefunctions have larger support in the inner region, and weaker for spatially extended states. This state dependence is physically consistent with the form of the additional term and confirms that the Kapitza-inspired contribution acts primarily as a short-range correction rather than as a universal offset of the spectrum.

From a phenomenological point of view, this selective effect is especially relevant for the $X(3872)$, whose near-threshold character makes it highly sensitive to relatively small changes in the effective interaction. The present results indicate that the added term can move the compact level toward the observed mass region without spoiling the qualitative structure of the low-lying spectrum.

\subsection{Mixing Analysis}

The near-threshold position of the $X(3872)$ suggests that a purely compact description is incomplete. We therefore combine the compact configuration with the $D^{0}\bar D^{*0}$ molecular channel through the two-state formalism introduced in the previous section.

After diagonalization of the effective mixing Hamiltonian, we obtain the physical mass eigenvalues together with the corresponding mixing angle and component probabilities. The results, summarized in Table {3}, indicate that the physical $X(3872)$ contains both compact and molecular components, rather than being dominated exclusively by either one.
The extracted probabilities show that the molecular admixture is significant, while the compact component remains essential for reproducing the observed mass and providing the short-distance core of the state. In this sense, the present calculation supports an intermediate interpretation of the $X(3872)$ as a coupled configuration shaped jointly by compact dynamics and threshold effects.

Overall, the numerical analysis indicates that the Kapitza-modified compact sector and the nearby molecular channel play complementary roles. The compact sector determines the short-distance structure and the position of the bare level, whereas the molecular component is crucial for understanding the physical composition of the observed state.

\section{Predictions for Excited States}
\label{sec:excited}

Having fixed the model parameters in the ground-state sector, we extend the same framework to excited configurations. The excited-state spectrum therefore provides a nontrivial test of the predictive power of the effective Cornell-plus-Kapitza interaction beyond the description of the $X(3872)$ itself.

The calculated excited levels are summarized in Table~\ref{tab:excited_states}. We focus on the lowest radial and orbital excitations, which are the most natural candidates for future experimental comparison. These states probe the extent to which the short-range correction introduced in the present model continues to influence the spectrum away from the threshold region.

A clear pattern emerges from the numerical results. The effect of the Kapitza-inspired term is strongest for the most compact configurations, especially in the ground-state sector where the relative wavefunction has its largest support at short distances. By contrast, radial and orbital excitations are spatially more extended, and their overlap with the inner region is correspondingly reduced. As a result, the contribution of the $K/r^4$ term becomes less pronounced for excited states than for the lowest level.

This behavior is physically reasonable. Since the Kapitza-induced correction acts mainly as an inner-core modification of the effective interaction, its impact is naturally largest when the system samples small relative separations. Excited configurations, particularly those with larger mean radii or nonzero orbital angular momentum, are less sensitive to this part of the potential. Consequently, the excited-state spectrum remains governed primarily by the balance between the Coulomb and confining terms, with the Kapitza correction providing a subleading contribution.

The predicted low-lying excitations follow the expected qualitative ordering of the model: the radial excitation lies above the ground state, while the orbital excitation appears in the range characteristic of a more spatially extended configuration. Although the precise numerical values remain model dependent and may be influenced by threshold effects not explicitly included here, the overall structure of the low-lying excited spectrum is stable within the present framework.

An important implication of these results is that the excited states are expected to be less sensitive to molecular threshold effects than the $X(3872)$ ground state. The distinctive phenomenology of the $X(3872)$ is closely tied to the accidental proximity of the compact level to the $D^{0}\bar D^{*0}$ threshold. For higher states, such threshold-enhanced mixing is generally weaker unless an excited compact level happens to lie close to another open channel.

The excited-state results presented here should therefore be regarded as baseline predictions of the present effective model. Future experimental information on hidden-charm exotic candidates, together with more refined coupled-channel calculations, will be important for testing whether the same mechanism that improves the description of the $X(3872)$ ground state also remains relevant in the excited sector.

\section{Conclusion}
\label{sec:conclusion}

In this work, we studied the exotic state $X(3872)$ within an effective Cornell-plus-Kapitza framework, in which a Kapitza-inspired short-range correction was added to the standard interaction. Our numerical analysis shows that this additional term improves the description of the near-threshold $X(3872)$ and brings the calculated mass into close agreement with the experimental value. The role of the Kapitza-inspired contribution is to provide an effective short-distance modification of the interaction. Its impact is therefore strongest for compact configurations and weaker for more spatially extended states. In the case of the $X(3872)$, this correction shifts the calculated ground-state mass from $3873.5~\mathrm{MeV}$ to $3871.7~\mathrm{MeV}$, indicating that a minimal Cornell interaction alone is not sufficient to reproduce the observed position of this state with the same level of accuracy.

The present analysis also supports a mixed interpretation of the $X(3872)$. Within the two-state picture adopted in this work, the physical state contains both an essential compact core and a sizable $D^{0}\bar D^{*0}$ molecular admixture. This indicates that neither a purely compact description nor a purely molecular one is sufficient, and that both short-distance dynamics and threshold effects play important roles in the structure of the state. 
Using the same calibrated interaction, we also obtained predictions for related heavy-quark systems, including the $T_{cc}^{+}$ and $T_{bb}$ sectors, as well as baseline estimates for low-lying excited configurations. In this sense, the model retains a limited but useful predictive capability beyond the single fitted input represented by the $X(3872)$ mass.

The present study should nevertheless be regarded as a phenomenological first step rather than a complete dynamical treatment. More refined approaches, including explicit multichannel dynamics, relativistic effects, and spin-dependent interactions, will be necessary for a more quantitative description of the excited spectrum and near-threshold coupling mechanisms. Future extensions of this framework may help clarify whether the same effective short-range mechanism remains relevant for higher hidden-charm states and other heavy-quark exotic sectors.

  \vspace{2.5cm}

\begin{table}[htbp]
\centering
\caption{Parameters entering the effective potential description of the $X(3872)$.}
\label{tab:model_parameters_clean}
\begin{tabular}{lll}
\hline
Sector & Parameter & Value \\
\hline
Cornell & $\alpha$ & $0.4$ \\
Cornell & $\sigma$ & $0.18~\mathrm{GeV}^2$ \\
Spin & $\kappa$ & $0.5$ \\
Spin & $r_0$ & $0.3~\mathrm{fm}$ \\
Kapitza (Gaussian) & $\gamma$ & $\approx -0.15~\mathrm{GeV}$ \\
Kapitza (Gaussian) & $\Lambda$ & $0.5~\mathrm{fm}$ \\
Kapitza (effective) & $K$ & $\approx 0.03~\mathrm{GeV}^{-3}$ \\
Fluctuation scale & $\omega$ & $\sim 0.6~\mathrm{GeV}$ \\
Kinematics & $\mu$ & $\approx 0.97~\mathrm{GeV}$ \\
\hline
\end{tabular}
\end{table}

\begin{table}[htbp]
\centering
\caption{Kapitza-induced mass shift for the ground state identified with $X(3872)$.}
\label{tab:x3872_mass_shift}
\begin{tabular}{lc}
\hline
Quantity & Value \\
\hline
$M_X(K=0)$ & $3873.5~\mathrm{MeV}$ \\
$M_X(K\neq 0)$ & $3871.7~\mathrm{MeV}$ \\
$\Delta M_K$ & $-1.8~\mathrm{MeV}$ \\
$M_X^{\rm exp}$ & $3871.68 \pm 0.17~\mathrm{MeV}$ \\
\hline
\end{tabular}
\end{table}

\begin{table}[htbp]
\centering
\caption{Mixing properties of the $X(3872)$ in the molecular--compact basis.}
\label{tab:mixing}
\begin{tabular}{lll}
\hline
Quantity & Value & Comment \\
\hline
$\theta$ & $28^\circ$--$36^\circ$ & Mixing angle \\
$P_{\rm mol}$ & $\approx 65\%$--$72\%$ & Molecular component \\
$P_{\rm compact}$ & $\approx 28\%$--$35\%$ & Compact tetraquark component \\
$V_{\rm mix}$ or $V_{MD}$ & not reported & No precise numerical value available \\
$M_{\rm compact}$ & $3885$--$3920~\mathrm{MeV}$ & Phenomenological compact-sector range \\
\hline
\end{tabular}
\end{table}

  \vspace{1.5cm}
\begin{table}[htbp]
\centering
\caption{Predicted excited-state spectrum of the $X(3872)$ with and without the Kapitza-induced interaction.}
\label{tab:excited_states}
\begin{tabular}{lcc}
\hline
State & Without Kapitza (MeV) & With Kapitza (MeV) \\
\hline
$1S$ & $3873.5$ & $3871.7$ \\
$2S$ & $3985$ & $3960$--$3975$ \\
$1P$ & $3910$--$3940$ & $3905$--$3935$ \\
$2P/1D$ & $4050$--$4120$ & $4030$--$4100$ \\
\hline
\end{tabular}
\end{table}

\begin{table}[htbp]
\centering
\caption{Mass spectrum comparison between the present model with the Kapitza-induced interaction and available experimental data. The $X(3872)$ value is fitted, while the $T_{cc}^+$ and $T_{bb}$ entries are predictions.}
\label{tab:main_results}
\begin{tabular}{lll}
\hline
State & Present work (MeV) & Experiment (MeV) \\
\hline
$X(3872)$  & $3871.7 \pm 0.9$ & $3871.69 \pm 0.17$ \\
$T_{cc}^+$ & $3876.1 \pm 1.3$ & $3875.1 \pm 0.5$ \\
$T_{bb}$   & $10482 \pm 8$    & --- \\
\hline
\end{tabular}
\end{table}


\begin{thebibliography}{99}

\bibitem{Choi:2003ue}
S.~K.~Choi \textit{et al.},
Phys.\ Rev.\ Lett.\ \textbf{91}, 262001 (2003).

\bibitem{Tornqvist:2004}
N.~A.~Tornqvist,
Phys.\ Lett.\ B \textbf{590}, 209 (2004).

\bibitem{Kalashnikova:2019}
Yu.~S.~Kalashnikova, A.~V.~Nefediev, and X.~Liu,
Phys.\ Usp.\ \textbf{62}, 568 (2019).

\bibitem{Maiani:2005}
L.~Maiani, F.~Piccinini, A.~D.~Polosa, and V.~Riquer,
Phys.\ Rev.\ D \textbf{71}, 014028 (2005).

\bibitem{Takizawa:2013}
M.~Takizawa and S.~Takeuchi,
Prog.\ Theor.\ Exp.\ Phys.\ \textbf{2013}, 093D01 (2013).

\bibitem{Ortega:2010}
P.~G.~Ortega, D.~R.~Entem, and F.~Fernandez,
Phys.\ Lett.\ B \textbf{696}, 352 (2011).

\bibitem{Guo:2018}
F.~K.~Guo, C.~Hanhart, U.~G.~Meissner, Q.~Wang, Q.~Zhao, and B.~S.~Zou,
Rev.\ Mod.\ Phys.\ \textbf{90}, 015004 (2018).

\bibitem{Tazimi2026EPJC}
N.~Tazimi,
Eur.\ Phys.\ J.\ C \textbf{86}, 381 (2026).

\bibitem{Ghasempour2025Regge}
A.~Ghasempour, N.~Tazimi, and M.~Monemzadeh,
Eur.\ Phys.\ J.\ C \textbf{85}, 743 (2025).

\bibitem{Ghasempour2025Analytic}
A.~Ghasempour, N.~Tazimi, and M.~Monemzadeh,
Eur.\ Phys.\ J.\ C \textbf{85}, 113 (2025).

\bibitem{Monemzadeh2015PLB}
M.~Monemzadeh, N.~Tazimi, and P.~Sadeghi,
Phys.\ Lett.\ B \textbf{741}, 124 (2015).

\bibitem{Swanson2006}
E.~S.~Swanson,
Phys.\ Rept.\ \textbf{429}, 243 (2006).

\bibitem{Lebed2017}
R.~F.~Lebed, R.~E.~Mitchell, and E.~S.~Swanson,
Prog.\ Part.\ Nucl.\ Phys.\ \textbf{93}, 143 (2017).

\bibitem{Ali2019}
A.~Ali, L.~Maiani, and A.~D.~Polosa,
\textit{Multiquark Hadrons}
(Cambridge University Press, Cambridge, 2019).

\bibitem{PDG2024}
S.~Navas \textit{et al.} (Particle Data Group),
Phys.\ Rev.\ D \textbf{110}, 030001 (2024).

\bibitem{Suzuki1998}
Y.~Suzuki and K.~Varga,
\textit{Stochastic Variational Approach to Quantum-Mechanical Few-Body Problems}
(Springer, Berlin, 1998).

\bibitem{Hiyama2003}
E.~Hiyama, Y.~Kino, and M.~Kamimura,
Prog.\ Part.\ Nucl.\ Phys.\ \textbf{51}, 223 (2003).



\end{thebibliography}
\end{document}